# High Open-Circuit Voltages in Lead-Halide Perovskite Solar Cells – Experiment, Theory and Open Questions

Thomas Kirchartz[1,2*]
[1]IEK5-Photovoltaik, Forschungszentrum Jülich, 52425 Jülich, Germany
[2]Faculty of Engineering and CENIDE, University of Duisburg-Essen, Carl-Benz-Str. 199, 47057 Duisburg, Germany

## Abstract

One of the most significant features of lead-halide perovskites are their ability to have comparably slow recombination despite the fact that these materials are mostly processed from solution at room temperature. The slow recombination allows achieving high open-circuit voltages when the lead-halide perovskite layers are used in solar cells. This perspective discusses the state of the art of our understanding and of experimental data with regard to recombination and open-circuit voltages in lead-halide perovskites. A special focus is put onto open questions that the community has to tackle to design future photovoltaic and optoelectronic devices based on lead-halide perovskites and other semiconductors with similar properties.

# 1. Introduction

During most of the history of photovoltaic power conversion, there has been a clear correlation between the amount of effort and energy that was put into the preparation of a light absorbing semiconductor and the electronic quality of said semiconductor. The longest charge-carrier lifetimes were correlated with the lowest defect densities and therefore usually the highest crystallinity of the bulk material and the interfaces. If we look at the top of the solar cell efficiency charts [1], this is still true given that epitaxially grown GaAs and monocrystalline Si solar cells [2,3] have the highest efficiencies of all single junction solar cells as shown in Fig. 1. What has changed is the ranking among the thin-film technologies that were dominated by Cu(In,Ga)Se$_2$ [4-6] and CdTe at least from the perspective of record efficiency on the lab scale. Since 2018, the non-stabilized efficiency records for single junction thin-film solar cells are held by lead-halide perovskite solar cells [7], despite the fact that these are solution processed semiconductors and do not even possess an optimal band gap for single junction solar cells.

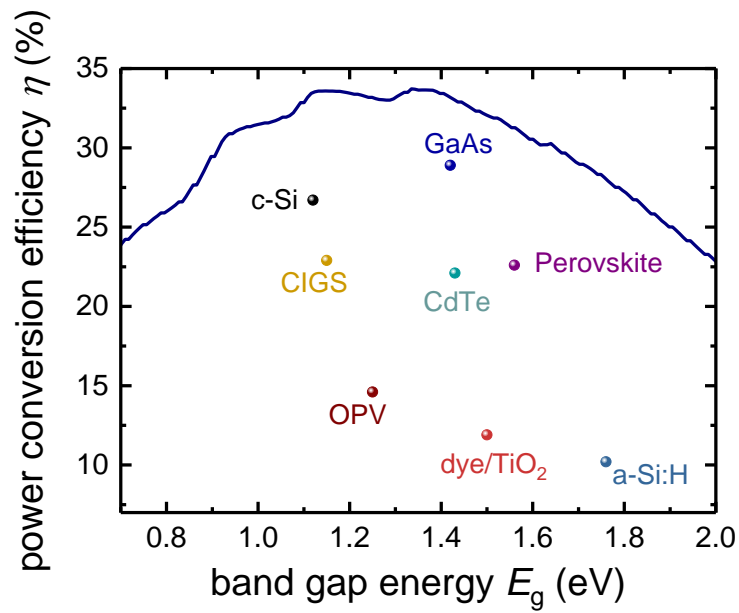

**Figure 1**: Power conversion efficiency in the Shockley-Queisser model (solid line) as a function of band gap compared with experimentally achieved values of efficiency in different photovoltaic technologies. Data are taken from the NREL efficiency chart [8], the efficiency tables [1] and some individual papers [7,9] in cases where the record in the references above

did not have associated data that allowed estimating the band gap of this particular compound or molecular blend. In case of the perovskite solar cells, the current efficiency record according to the NREL efficiency chart is 23.7 %. The best certified and published data from ref. [7] is 22.6 %. (Online version in colour.)

The key requirement for achieving efficient photovoltaic power conversion is to combine high absorption coefficients with long charge-carrier lifetime.[10,11] An obvious problem for assigning a simple figure of merit to this insight is that the absorption coefficient is a function of energy, so the magnitude and shape of the absorption coefficient matters. For instance, the high absorption coefficient of Si above its direct band gap at 3.4 eV is clearly much less relevant for photovoltaics than its low absorption coefficient close to the indirect band gap at 1.12 eV. However, when studying only direct semiconductors, there are relatively few differences in the shape and sharpness of the absorption between the most important photovoltaic materials [12]. Thus, it is the charge-carrier lifetime that differs between the different technologies and that makes the difference between high and low power conversion efficiencies [13]. In the following, we will therefore discuss the concept of the charge-carrier lifetime in more detail. In halide perovskites just as in other direct semiconductors [14], charge carrier lifetimes are often studied using the method of transient photoluminescence [15-19]. We therefore show some exemplary transient photoluminescence data on halide perovskite layers on a hole transport layer and discuss how to interpret and analyze the data in terms of charge-carrier lifetimes and surface recombination velocities. Subsequently, I will discuss the step from high carrier lifetimes to high open-circuit voltages and completed solar cells. In order to rationalize the results, non-radiative recombination and the theory of multiphonon recombination is briefly reviewed before the key open questions in the context of recombination and high open circuit voltages in lead-halide perovskites are discussed.

## 2. Charge-Carrier Lifetimes

The charge-carrier lifetimes of semiconductor films, layer stacks or devices are both highly relevant for its functionality as well as subject to a variety of more or less intuitive definitions that complicate an understanding of their meaning. The charge-carrier lifetime describes the dependence of the electron or hole concentration as a function of time after the generation of electrons and holes (e.g. by a laser pulse) has stopped. In a doped semiconductor, at low excitation densities, one type of carrier will be in excess of the other. This situation is termed low-level injection. For instance in a *p*-type semiconductor, the electrons would be the minority carriers while the hole concentration $p$ would be constant as a function of time $t$ and given by the density $N_A$ of ionized acceptors. Clearly, in this situation, it only makes sense to define an *electron* lifetime, while the hole lifetime would be infinite because it does not decay at all with time. In contrast, if the excitation density is high enough and the doping density low enough that excess charge carriers of both types exceed the concentration of carriers due to ionized dopants, the sample is in high-level injection and there will be a finite lifetime for both types of carriers.

In addition to the aspect of injection level, lifetimes can be associated with different recombination mechanisms, such as radiative recombination (with rate $R_{rad}$), Shockley-Read-Hall (SRH) recombination [20,21] via defects in the bulk (with rate $R_{SRH}$), SRH recombination at surfaces and Auger recombination (with rate $R_{Aug}$). Including the three bulk recombination mechanisms, the electron density $n$ as a function of time $t$ and position $x$ after a laser pulse is given by the differential equation

$$-\frac{dn}{dt} = R_{rad} + R_{SRH} + R_{Aug}$$
$$= \left[ k_{rad}(1-p_r) + \frac{1}{n\tau_p + p\tau_n} + C_n n + C_p p \right](np - n_i^2), \quad (1)$$

where $p$ is the hole density, $n_i$ the intrinsic carrier concentration, $k_{rad}$ is the radiative recombination coefficient, $p_r$ is the reabsorption probability taking into account the effect of photon recycling [22-25], $\tau_{n,p}$ are the SRH lifetimes for electrons and holes, and $C_{n,p}$ are the Auger coefficients for electrons and holes. This differential equation has boundary conditions for the surfaces of the semiconductor that consider the loss of electrons via surface recombination. The boundary condition at the edges $x = 0$ and $x = d$, where $d$ is the thickness, are given by

$$\pm qD_n \frac{dn}{dx}\bigg|_{x=0,d} = \frac{np - n_i^2}{n/S_p + p/S_n}\bigg|_{x=0,d}, \qquad (2)$$

where the sign on the left hand side is positive for $x = 0$ and negative for $x = d$. Here, $D_n$ is the diffusion constant for electrons and $S_n$ and $S_p$ are the surface recombination velocities for electrons and holes. Note that equations analogous to Eq. (1) and (2) can be written down for the case of holes. Equations (1) and (2) imply that the measured decay may be influenced by four different recombination mechanisms (radiative, bulk defect, surface defect and Auger) and that in general each of the different mechanisms is a non-linear function of $n$ or $p$. Equations (1) and (2) alone do not immediately suggest that the concept of a lifetime is suitable to describe the decay of electrons or holes after a pulse. However, the concept of a lifetime becomes substantially more intuitive and meaningful in the case of low level injection that is relevant for many inorganic semiconductors. In low level injection, each of the recombination terms can be simplified such that it depends linearly on minority-carrier density. For instance in a $p$-type semiconductor, eq. (1) becomes

$$-\frac{dn}{dt} = \left[k_{rad}(1-p_r)N_A + \frac{1}{\tau_n} + C_p N_A^2\right](n-n_0)$$
$$= \frac{(n-n_0)}{\tau_{eff}}, \qquad (3)$$

where the effective lifetime $\tau_{eff}$ is defined as the inverse of the square brackets in Eq. (3). Even the inclusion of surfaces in the concept of an effective lifetime is possible for low level injection. For the case that the surface recombination velocities for electrons as minorities are the same at both surfaces, a surface lifetime can be determined as [26]

$$\tau_s = \frac{d}{2S_n} + \frac{d^2}{D_n \pi^2}, \qquad (4)$$

Here, the first term on the right hand side describes the recombination at the surface itself and the second term describes the diffusion to the surface which may be limiting for large thicknesses $d$, low diffusion constants $D_n$ and high surface recombination velocities $S_n$. The total effective lifetime in low level injection would then follow from adding up the inverse lifetimes and then inverting the resulting sum again leading to

$$\tau_{eff} = \left[ k_{rad}(1-p_r)N_A + \frac{1}{\tau_n} + C_p N_A^2 + \frac{2S_n D_n \pi^2}{d(D_n \pi^2 + 2S_n d)} \right]^{-1}. \qquad (5)$$

Note that $\tau_{eff}$ in Eq. (5) is a constant, carrier concentration independent value. This is due to the fact that even the two recombination mechanisms (radiative and Auger) that are in general non-linear, become linear in minority carrier concentration in low level injection. The doping density sets a natural upper limit to the effective lifetime, which can never exceed the value in the radiative limit that is given by $\tau_{eff} = \left[ k_{rad}(1-p_r)N_A \right]^{-1}$. Because $k_{rad}$ in direct semiconductors is typically on the order of $10^{-10}$ cm$^3$/s (e.g. $2\times10^{-10}$ cm$^3$/s in GaAs [27,28], $0.6\times10^{-10}$ cm$^3$/s to $2\times10^{-10}$ cm$^3$/s in Cu(In,Ga)Se$_2$ with different In to Ga ratios [29]), lifetimes of hundreds of ns are already a good value for direct semiconductors at typical doping concentrations of $10^{16}$ cm$^{-3}$. Typical values reported for the radiative and bimolecular recombination coefficients in the literature for CH$_3$NH$_3$PbI$_3$ are presented in table I. In contrast, in indirect semiconductors such as Si, $k_{rad}$ is typically substantially smaller (in Si $k_{rad}$ = $4.7 \times 10^{-15}$ cm$^3$/s [30]) allowing much longer lifetimes. For instance, in monocrystalline Si with good surface passivation, lifetimes of tens of milliseconds are possible [31]. Such lifetimes could (theoretically) be achieved in GaAs only for extremely low doping densities of order $10^{12}$ cm$^{-3}$.

**Table I:** Values for the bimolecular recombination coefficients in units of cm$^3$/s of CH$_3$NH$_3$PbI$_3$ from literature taken from ref. [32]. Here, $k_{ext}$ is the externally measured bimolecular recombination coefficient, $k_{rad}$ is the radiative recombination coefficient obtained

after correcting for photon recycling and $k_{non}$ is any non-radiative contribution to the bimolecular recombination coefficient observed. In case of the data from Crothers et al. [33], the range for $k_{ext}$ is caused by a range of sample thicknesses that lead to a unique value of $k_{rad}$ if photon recycling is taken into account.

| Reference | $k_{ext}$ (cm$^3$/s) | $k_{rad}$ (cm$^3$/s) | $k_{non}$ (cm$^3$/s) |
|---|---|---|---|
| Staub et al. [23] | $4.78 \times 10^{-11}$ | $8.7 \times 10^{-10}$ | neglected |
| Staub et al. [23,24] | $4.78 \times 10^{-11}$ | $8.4 \times 10^{-11}$ | $4.4 \times 10^{-11}$ |
| Crothers et al. [33] | $1.4 \times 10^{-10}$ to $2 \times 10^{-11}$ | $6.8 \times 10^{-10}$ | 0 |
| Richter et al. [34][c] | $8.1 \times 10^{-11}$ | $7.1 \times 10^{-11}$ | $7.2 \times 10^{-11}$ |
| Richter et al. [34][d] | $7.9 \times 10^{-11}$ | $1.8 \times 10^{-10}$ | $5.6 \times 10^{-11}$ |

For lead-iodide perovskite films, layer stacks and devices, transient measurements such as transient photoluminescence are typically done in high-level injection due to the low doping density of the materials. This implies that both radiative and Auger recombination will lead to terms in the recombination rate that are quadratic or cubic in charge carrier density thereby leading to non-exponential decays [23]. Figure 2 shows a photoluminescence decay of a methylammonium-lead-iodide (MAPI) film deposited on a glass/ITO substrate coated with a thin layer of PTAA (poly(triarylamin)) that is used as a hole contact in solar cells and light emitting diodes [35]. We note that the decay is faster for short times followed by a nearly exponential decay at longer times. The fast features at short times can be explained by radiative recombination. At longer times, the long decay can either be due to bulk recombination or surface recombination at the MAPI/PTAA interface. In addition, there a range of additional effects such as lateral diffusion away from the illuminated spot, charge transfer to the contact layers, charge accumulation in the contact layers [36] and lateral transport on the contact layers that may affect the shape of the transients. While the two effects cannot be distinguished unambiguously, it is possible to perform numerical simulations [35,36] to include the different effects mentioned above and to estimate minimum values for SRH lifetime in the bulk and upper limits for the surface recombination velocity. In the present case, Fig. 2 suggests that $S < 1$ cm/s and $\tau_{bulk} > 5$ µs best describe the experimental decay. To put the values of surface recombination velocities into perspective, a brief word on typical values for inorganic semiconductors used for photovoltaics. In crystalline Si wafers, good surface passivation leads to surface recombination velocities just below 10 cm/s [31]. In the case of AlGaAs/GaAS or GaInP/GaAs interfaces, surface recombination velocities on the order of 1 to 15 cm/s have been reported already in the 1980s [37,38]. In CdS/CdTe heterstructures surface recombination velocities are typically on the order of $10^5$ cm/s [39] while high quality MgCdTe/CdTe heterostructures have surface recombination velocities on the order of 1 cm/s [40].

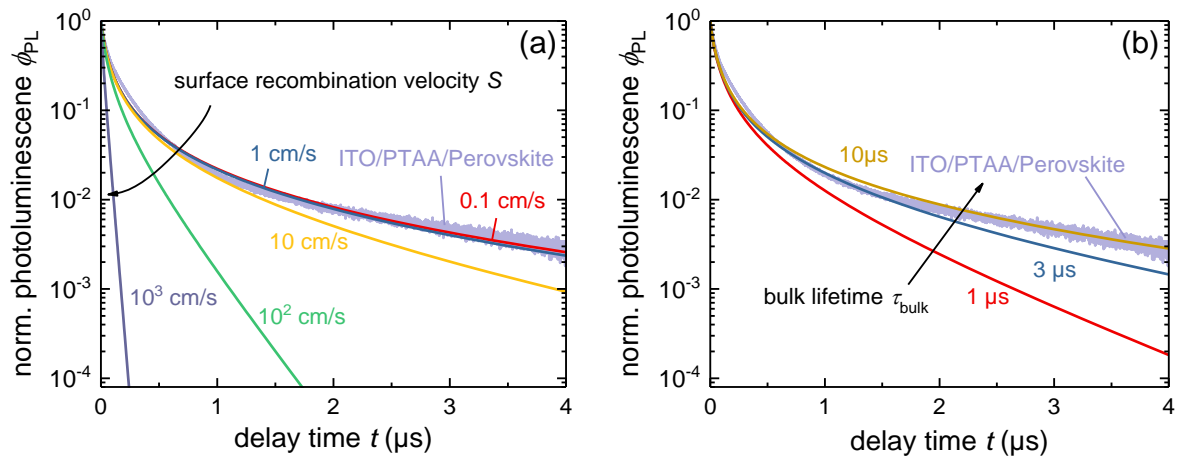

**Figure 2**: Experimental transient photoluminescence data of a MAPI film on a glass/ITO/PTAA substrate compared with numerical simulations varying (a) the surface recombination velocity (with $\tau_{bulk} = 10$µs = const) at the PTAA/MAPI interface and (b) the bulk lifetime (with $S = 0$). This comparison does not yield precise values for either $S$ or $\tau_{bulk}$ but it allows estimating lower limits for $\tau_{bulk}$ and upper limits for $S$. Data is redrawn from ref. [35]. (Online version in colour.)

## 3. Realizing High Open-Circuit Voltages in Solar Cells

We have shown in the previous chapter that long bulk lifetimes and low surface recombination velocities $S < 10$ cm/s are possible in perovskite layers and at perovskite/PTAA interfaces. In order to actually benefit from these promising values in a complete device, it is necessary to achieve low surface recombination velocities at both interfaces in the device and to maintain a good compromise between low surface recombination velocities, low series resistances of the used contact layers and a beneficial electric field distribution inside the absorber layer that optimizes charge collection and minimizes recombination at the maximum power point [41-45].

Figure 3 shows how for a constant thickness ($d$ = 300 nm) and a perfect defect free bulk, the open-circuit voltage, the voltage at the maximum power point and the power conversion efficiency of a MAPI solar cell depends on surface recombination and series resistance. The simulation includes the effects of Auger recombination using an Auger coefficient $C = 5.4 \times 10^{-28}$ cm$^6$/s [46] and photon recycling as described in more detail in [47-49]. We note that for MAPI efficiencies > 25 % should be possible but require series resistances below 5 $\Omega$cm$^2$ and surface recombination velocities below 100 cm/s at both contacts. The maximum possible open-circuit voltage is 1.32 eV for MAPI with a band gap $E_g$ = 1.6 eV and it decreases for $S$ > 100 cm/s, consistent with recent reports by Wang et al. [50]. Note that the exact dependence of $V_{oc}$ on $S$ also depends on the minority carrier concentration at the two contacts which are determined in turn by the differences between contact workfunction and conduction band (at the cathode) or valence band (at the anode) of the perovskite. Here we use an offset of 0.1 eV on each side which leads to a built-in voltage $V_{bi}$ = 1.4 eV (band gap minus twice the offset of 0.1 eV). The effect of the (area-related) series resistance $R_s$ on power conversion efficiency $\eta$ and the voltage $V_{mpp}$ at the maximum power point is quite substantial and relatively linear with a doubling of $R_s$ leading to double the loss in $V_{mpp}$ relative to the optimum value for $R_s$ = 0 which is consistent with analytical equations [51] describing the effect of $R_s$ on $V_{mpp}$. Resistive effects in perovskite solar cells may originate from external contact layers such as the ITO [52] but also from electron and hole contact layers, especially if these are made up of undoped organic semiconductors such as PTAA and PCBM or C$_{60}$ as is commonly done in *p-i-n*-type device stacks. The low permittivity of these electron and hole contact materials combined with the high permittivity [53] of the perovskite absorber layer and the electric field screening by ion movement may also lead to low electric fields in the perovskite absorber that may even change sign and impede majority carrier extraction as illustrated in refs. [54,55]. In these cases, the good passivation provided by at least some organic contact materials may compromise efficient charge extraction [43,44].

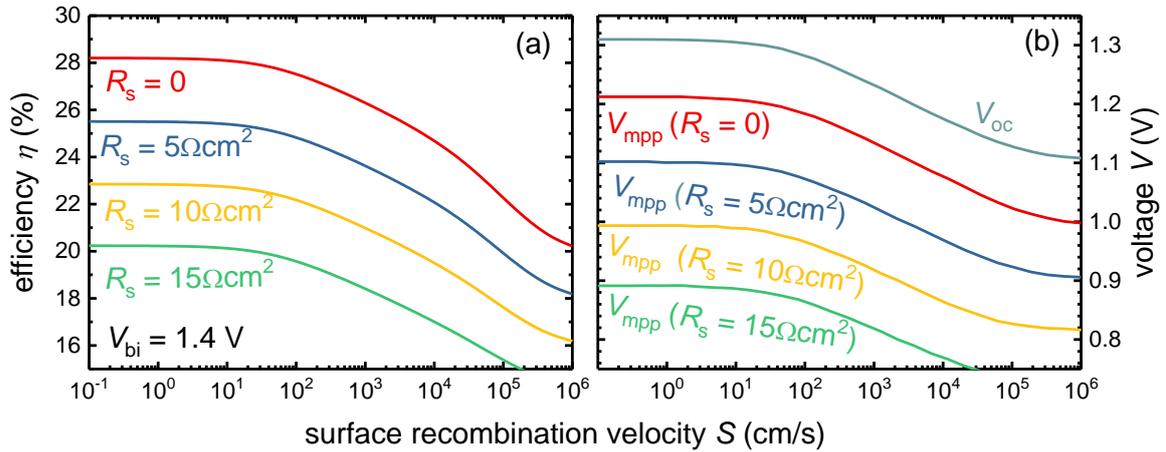

**Figure 3**: Result of drift-diffusion simulations of MAPI solar cells showing (a) power conversion efficiency and (b) open-circuit voltage $V_{oc}$ as well as the voltage $V_{mpp}$ at the maximum power point as a function of the surface recombination velocity $S$ for minority carriers at front and back contact. In addition to $S$, also the series resistance $R_s$ is varied with higher values of $R_s$ leading to lower efficiencies and voltages $V_{mpp}$. The open-circuit voltage stays constant with $R_s$ because there is no current flow in open circuit and therefore no voltage drop over the series resistance. While low values of $S$ guarantee high open-circuit voltages, what counts for efficiency is $V_{mpp}$ which may be reduced drastically by higher series resistances. This illustrates that contact layers not only have to ensure low $S$ but also low $R_s$. (Online version in colour.)

It is worth comparing the simulated open-circuit voltages shown in Fig. 3b with experimentally achieved values of the open-circuit voltage $V_{oc}$. Figure 4a shows $J_{sc}$ as a function of $V_{oc}$ for a range of highly efficient lead-halide perovskite solar cells reported in literature (see table II for a full set of values and references). The solid line defines the respective values of $J_{sc}$ and $V_{oc}$ in the Shockley-Queisser model. Figure 4a shows that typical values for most of the high efficiency lead-halide perovskite cells are $V_{oc} \approx 1.15$ V and $J_{sc} \approx$ 22 mA/cm². Some are optimized for higher open-circuit voltages $V_{oc} > 1.2$ V [15,35] and some have slightly lower band gaps allowing higher photocurrents around $J_{sc} \approx 25$ mA/cm² [7,56].

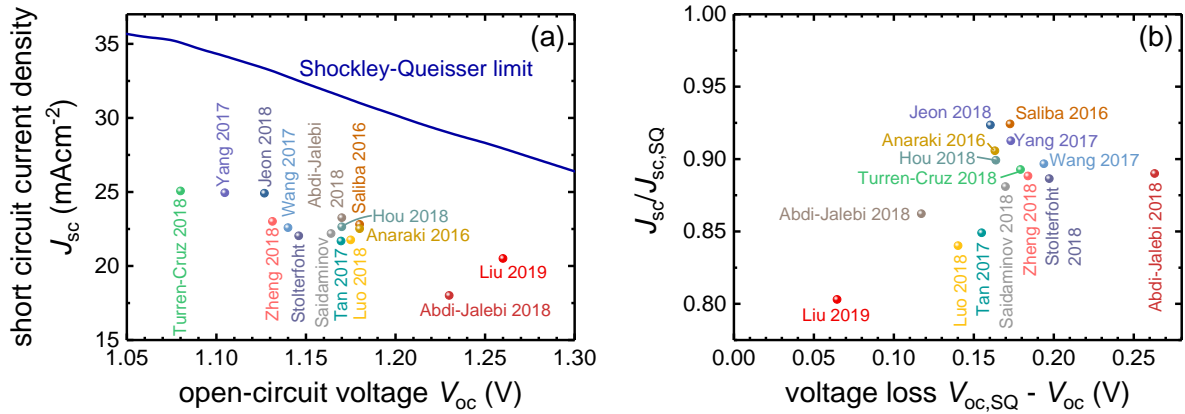

**Figure 4**: (a) Short-circuit current density $J_{sc}$ vs. open-circuit voltage $V_{oc}$ for a range of lead-halide perovskite solar cells with high efficiencies (symbols) as well as the maximum values for $J_{sc}$ and $V_{oc}$ given by the Shockley-Queisser model (solid line) [57]. (b) Now the values for $J_{sc}$ and $V_{oc}$ are presented relative to the values in the Shockley-Queisser model, which helps identifying the potential for improvement for the different devices. (Online version in colour.)

However, Figure 4a is not well suited to identify potential for improvement for a given device because the band gap varies slightly between the different perovskite compositions. Therefore, Figure 4b provides a more direct relation between the experimental data and the values given by the Shockley-Queisser model by normalizing $J_{sc}$ to $J_{sc,SQ}$ and taking the difference between the open-circuit voltage in the SQ limit and the real open-circuit voltage. Figure 4b shows that many of the high efficiency devices reach > 90 % of the photocurrent in the SQ model which is a good value compared to many other technologies. However, values of ~95% are possible in crystalline Si and CdTe solar cells [58,59], i.e. there is still some room for improvement. The voltage loss in open circuit is typically > 120 mV for most devices suggesting that there is indeed still some room for improvement given that GaAs solar cells have voltage losses < 40mV in the best cases (see e.g. Fig. 2b in ref. [60]).

**Table II:** Device performance data of a range of small area lead-halide perovskite solar cells used for Figure 4. The band gap is as reported in the papers or estimated from the external quantum efficiency data provided. The value $\Delta V_{oc}$ is defined as the difference between the $V_{oc}$ in the Shockley-Queisser model calculated based on the indicated band gap and the actual $V_{oc}$. Datasets with an asterisk are certified data.

| Reference | $E_g$ (eV) | $\eta$ | $J_{sc}$ (mAcm$^{-2}$) | $V_{oc}$ (V) | FF | $\Delta V_{oc}$ (V) | $J_{sc}/J_{sc,SQ}$ |
|---|---|---|---|---|---|---|---|
| Jeon 2018 [7]* | 1.56 | 22.6 | 24.9 | 1.13 | 80.5 | 0.160 | 0.924 |
| Luo 2018 [61]* | 1.59 | 20.9 | 21.8 | 1.18 | 81.7 | 0.140 | 0.840 |
| Turren-Cruz 2018 [56] | 1.53 | 20.4 | 25.1 | 1.08 | 75.5 | 0.179 | 0.893 |
| Stolterfoht 2018 [62]* | 1.62 | 19.6 | 22.0 | 1.15 | 77.6 | 0.197 | 0.886 |
| Saidaminov 2018 [63] | 1.61 | 21.0 | 22.2 | 1.16 | 81.3 | 0.170 | 0.881 |
| Zheng 2018 [64]* | 1.59 | 20.6 | 23.0 | 1.13 | 79.1 | 0.184 | 0.888 |
| Wang 2017 [65] | 1.61 | 20.6 | 22.6 | 1.14 | 80.0 | 0.194 | 0.897 |
| Saliba 2016 [66] | 1.63 | 21.8 | 22.8 | 1.18 | 81.0 | 0.173 | 0.924 |
| Tan 2017 [67]* | 1.60 | 20.1 | 21.7 | 1.17 | 79.4 | 0.155 | 0.849 |
| Yang 2017 [68]* | 1.55 | 22.1 | 24.9 | 1.10 | 80.3 | 0.173 | 0.913 |
| Abdi-Jalebi 2018 [15] | 1.56 | 21.5 | 23.3 | 1.17 | 79.0 | 0.117 | 0.862 |
| Abdi-Jalebi 2018 | 1.78 | 17.5 | 18.0 | 1.23 | 79.0 | 0.263 | 0.890 |
| Hou 2017 [69] | 1.61 | 21.2 | 22.6 | 1.17 | 80.0 | 0.164 | 0.899 |
| Anaraki 2016 [70] | 1.62 | 20.5 | 22.5 | 1.18 | 77.0 | 0.163 | 0.906 |
| Liu 2019 [35] | 1.60 | 20.7 | 20.5 | 1.26 | 80.1 | 0.065 | 0.803 |

Among the solar cells presented in Fig. 4 and table II, the device approaching the $V_{oc,SQ}$ most closely is the one labeled Liu 2019. It is based on a simple MAPI cell with a lead-acetate precursor with 8% lead-chloride and using ITO/PTAA and PCBM/BCP/Ag as hole and electron contact layers, respectively. Figure 5 shows the power density and current density curves of this MAPI solar cell (shown for the downscan at 0.1 V/s) and compares it with three computed scenarios. Each scenario keeps the parameters of the previous case but increases first the *FF*, then $V_{oc}$ and finally $J_{sc}$ to the ideal value that is defined by the SQ model. The differences between the curves, then allow us to study the losses due to each mechanism. The *FF* of the experimental curve is 80%, but it may be up to 90% for this open-circuit voltage [71,72]. This corresponds to an increase by about 120 mV in the voltage at the maximum power point. Further increasing $V_{oc}$ by removing all non-radiative recombination would only yield another ~60 mV in $V_{oc}$ and $V_{mpp}$. A further substantial gain in power conversion efficiency would be possible if more complete light absorption was achieved. Finally, increasing $J_{sc}$ to $J_{sc,SQ}$ would yield an efficiency of ~30 %, i.e. ~10% more than the experimental value and ~ 5% more than the value where $V_{oc}$ and *FF* are already taken as ideal. Thus, for this specific device, the open-circuit voltage is nearly fully optimized with little additional gains being expected. In contrast, the photocurrent is far from optimized, which sets optical means to trap light, processes to fabricate thicker layers, optimization of charge collection and minimization of parasitic absorption on the research agenda.

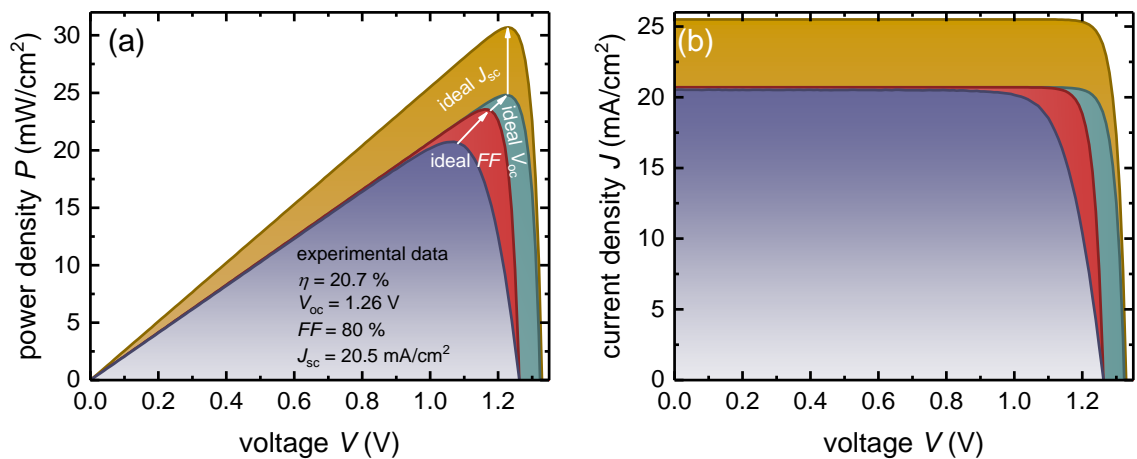

**Figure 5**: (a) Power density vs. voltage and (b) current density vs. voltage curve of a MAPI solar cell (data taken from ref. [35]) compared with more idealized scenarios where first the *FF* is taken as ideal, then the open-circuit voltage and finally the short-circuit current density. This loss analysis shows that for the current cell, the most substantial losses relative to the Shockley-Queisser limit occur in light absorption followed by resistive losses leading to lower fill factors. Non-radiative recombination leading to reduced open-circuit voltages is the smallest of the three loss terms. (Online version in colour.)

## 4. Multiphonon Recombination

We learned in the previous sections that solution processed lead-halide perovskites allow closely approaching the thermodynamic limit for the open-circuit voltage. This is possible thanks to very long bulk lifetimes of several microseconds and by achieving surface recombination velocities that are lower than those of the best passivation layers for crystalline Si solar cells [31]. Thus, the question remains how we can rationalize these long lifetimes. There have been various physical mechanisms suggested in the literature to explain slow recombination [73-79]. Some mechanisms such as the existence of an indirect gap [80,81] serve to explain slow radiative recombination. However, radiative recombination in MAPI is not actually slow but corresponds to what is expected [23,82] based on the absorption coefficient of MAPI and the fairly low effective densities of states. In addition, slow radiative recombination would not help achieving higher efficiencies as detailed in refs. [11,83]. Only slow non-radiative recombination would actually be helpful. In order to achieve slow non-radiative recombination a semiconductor (bulk or interface) needs to have both a low density of defects and low transition rates between at least one of the bands and the defect. If the defect only interacts efficiently with one of the bands, recombination will be less likely to happen. Instead the defects will trap and then later detrap charge carriers. While this mechanism slows down transport, it would not lead to reduced open-circuit voltages.

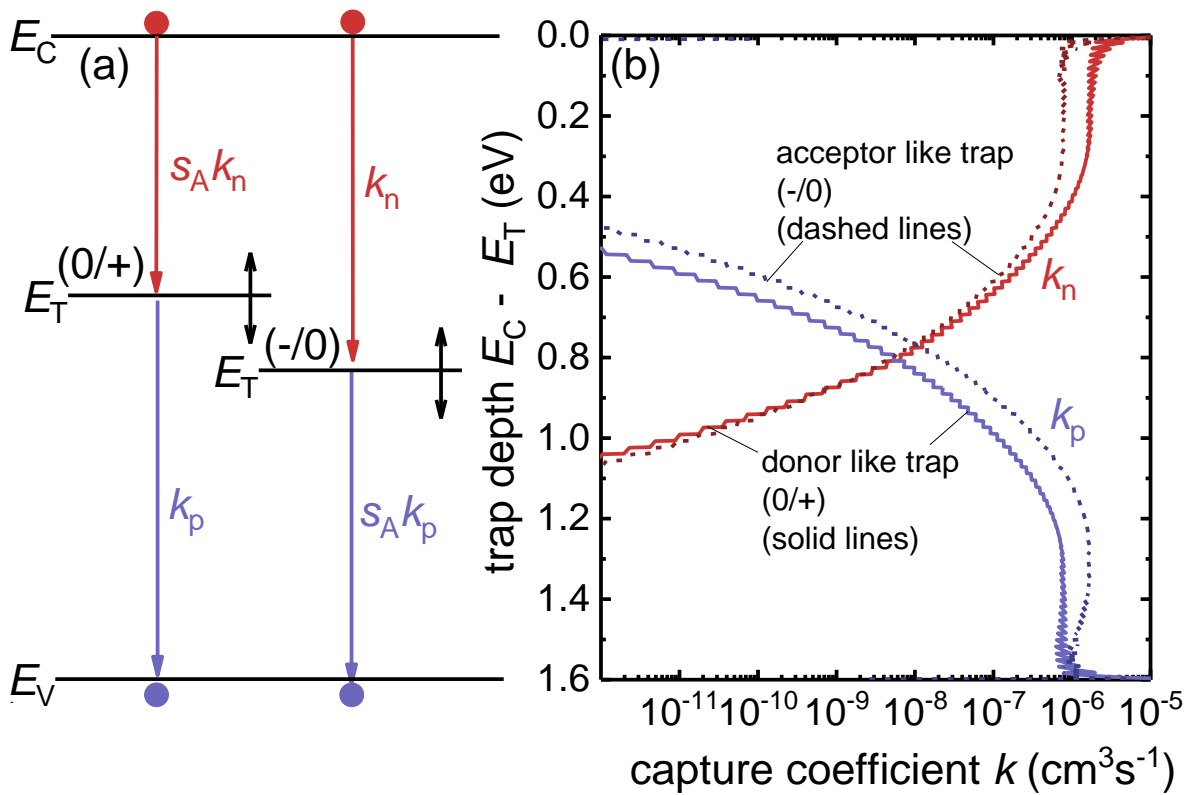

**Figure 6**: (a) Schematic of conduction band, valence band and a donor and acceptor like trap. The downward transitions have coefficients $k_{n,p}$, some of which are amplified by Coulomb attraction between charge carrier and charge state of the trap. these are the transitions between free electrons and positively charged (donor-like) trap and between free holes and a negatively charged (acceptor-like) trap. (b) Capture coefficients $k_{n,p}$ including the effect of Coulomb attraction as a function of the position of a singly charged (acceptor like or donor like) trap whose position may change between conduction and valence band edge as indicated in panel (a). The capture coefficients are calculated using semi-classical multiphonon theory and trap-depth and charge state dependent Huang-Rhys factors as discussed in refs. [84-86]. (Online version in colour.)

Thus, in high level injection, where $n \approx p$, defects with substantially asymmetric capture coefficients for electrons and holes might be less detrimental than those with symmetric capture coefficients. Asymmetry of capture coefficients may originate for instance from the position of the defect and from its charge state. For instance, defects close to the conduction band are more likely to interact with electrons in the conduction band than with holes in the valence band. In addition, positively charged defects are more likely to capture electrons than neutral defects. Thus, it would be useful to be able to estimate the relevance of each of these effects on the value of the capture coefficients. These can be calculated for non-radiative processes using the theory for multiphonon recombination, which assumes that every downward transition between two states has to involve the emission of multiple phonons in order to dissipate the energy of the excited state [87,88]. Figure 6 shows the capture coefficients calculated using the theory discussed in ref. [84] and originating from the work of Markvart [85,87] and Ridley [86,88]. The capture coefficients show a strong dependence on the position of the defect with shallow defects having one high and one extremely low capture coefficient while for deep defects the values of the capture coefficients intersect. When comparing the graph for the donor-like and the acceptor-like defect, we note that the effect of Coulomb attraction on the capture coefficients is fairly small.

In order to calculate the Shockley-Read-Hall lifetimes introduced in Eq. (1) from the capture coefficients in Fig. 7, we need to specify a trap density. The SRH lifetimes are given by $\tau_{n,p} = \left(k_{n,p} N_T\right)^{-1}$, where $N_T$ is the trap density. If we choose $N_T = 10^{15}$ cm$^{-3}$, we obtain the lifetimes shown in Fig. 7a using the capture coefficients $k_{n,p}$ shown in Fig. 6. From the SRH lifetimes, we can then calculate the effective lifetime $\tau_{eff}$, which in the limit of high level injection assumed here ($n \approx p$) is just given as $\tau_{eff} = \tau_n + \tau_p$.

The Coulomb attraction is calculated in ref. [84,89] by using the Sommerfeld factor to modify the capture cross sections of transitions with attractive Coulomb interaction. The Sommerfeld factor for attractive transitions is given by [89]

$$s_A = 4\sqrt{\frac{\pi m_{eff}}{\varepsilon_r^2 m_0}\frac{R_H}{kT}}. \qquad (6)$$

where $m_{eff}/m_0$ is the effective mass relative to the electron rest mass, $R_H \approx 13.6$ eV is the Rydberg energy, $kT$ is the thermal energy and $\varepsilon_r$ is the low frequency dielectric permittivity. Figure 7b illustrates the result of Eq. (6) for different values of $m_{eff}/m_0$ and $\varepsilon_r$. It becomes clear that for the case of low effective masses and high permittivities, the Sommerfeld factor does not deviate much from unity.

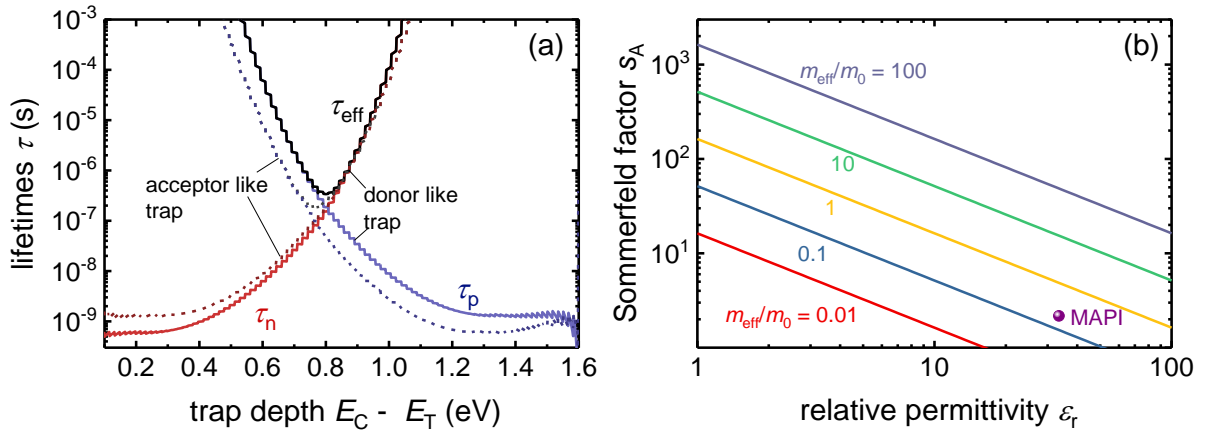

**Figure 7**: Shockley-Read Hall lifetimes $\tau_n$ and $\tau_p$ obtained using multiphonon theory as outlined in ref. [84] to obtain the capture coefficients for transitions between the band and localized trap states. The lifetimes are given as a function of the depth of the trap relative to the conduction band edge $E_C$ and for the case of a donor like (solid line) and an acceptor like defect. From $\tau_n$ and $\tau_p$ an effective lifetime is calculated assuming high level injection ($n \approx p$) via $\tau_{eff} = \tau_n + \tau_p$. (b) Sommerfeld factor $s_A$ for attractive transitions (e.g. electron is trapped by a positively charged defect) as a function of the low frequency permittivity $\varepsilon_r$ and for different values of the effective mass $m_{eff}/m_0$. Using typical values reported for MAPI ($\varepsilon_r = 33.5$ [53], $m_{eff}/m_0 = 0.2$), the Sommerfeld factor assumes the value ~2.2 indicated in the graph as a filled circle. This low value implies a weak influence of attractive vs. neutral defect states and therefore a weak shift of the data for donor and acceptor like defects shown in panel (a). (Online version in colour.)

Thus, the main contribution to the capture coefficients in the framework of multiphonon recombination using the parameters of MAPI as input would be the position of the trap.

Midgap traps would still yield reasonably high recombination coefficients in the framework of multiphonon recombination theory. However, already small shifts of the trap position relative to midgap would strongly reduce the efficiency of recombination. Density functional theory calculations in the bulk [90] and the surface [91] of MAPI have shown that most defects in MAPI are shallow with the iodine interstitial being the only intrinsic deep point defect. However, even the iodine interstitial is likely not a midgap defect. This finding in combination with Fig. 7a suggests that MAPI could tolerate quite high densities of intrinsic point defects given that they are not likely to have symmetric and high capture coefficients.

## 5. Open Questions

While the experimental results with respect to slow recombination and high open-circuit voltages are certainly impressive and while we have some understanding and rationale for explaining the experimental results, there remain a range of open questions that should be tackled in the future.

Recently, it has been shown that the model of the harmonic oscillator for the potential energy of electrons and holes is insufficient to explain the magnitude and temperature dependence of the charge carrier mobility in lead-halide perovskites [92]. Instead, the anharmonic fluctuations of the lead-halide bonds affect the overlap of wavefunctions and thereby charge transport. In a similar way also the multiphonon recombination models that rely on harmonic oscillators to calculate transition rates between band like states and defect states likely have to be modified to the special case of lead-halide perovskites.

In addition, the models for recombination generally only describe bulk recombination but not interface recombination. At an interface, defects are more likely to form than in the bulk. Therefore for most inorganic solar cell materials, interface recombination is even more important than bulk recombination. Recently, it has been argued [93-95] that the high energy of vibrational modes in organic semiconductors leads to fast recombination, especially for lower band gaps. This concept applied to interfaces between perovskite absorbers and organic electron and hole contact layers would suggest that organic interface materials might accelerate recombination because it is easier to dissipate the energy into the organic material. However, this doesn't seem to be the case, given that perovskite/organic interfaces have extremely low surface recombination velocities [35,62,96]. Thus, a better fundamental understanding of interface recombination is needed to *design* interfaces rather than find suitable ones by trial and error.

The low surface recombination velocities of organic electron and hole transport layers might bring about a fundamental compromise between recombination, resistive losses and optimization of the electric field distribution in the device [43,44]. As shown in Fig. 3, a low value of $S$ might benefit $V_{oc}$ but if it comes with a high value of $R_s$, the efficiency might be reduced. The same holds true if the electric field distribution is such that it makes charge extraction more difficult, because most of the electric field drops over the electron and hole transport layers [43,44]. Thus, we need a better understanding of resistive and charge collection losses at interfaces and in electron and hole transport layers. While it is generally easy to measure lumped series resistances in solar cells [97,98], it is more difficult to disentangle [99,100] the influence of different layers on the series resistance. Thus, methods are needed that measure how the internal quasi-Fermi level splitting in the perovskite absorber is coupled to the external quasi-Fermi level splitting at the contacts at the maximum power point.

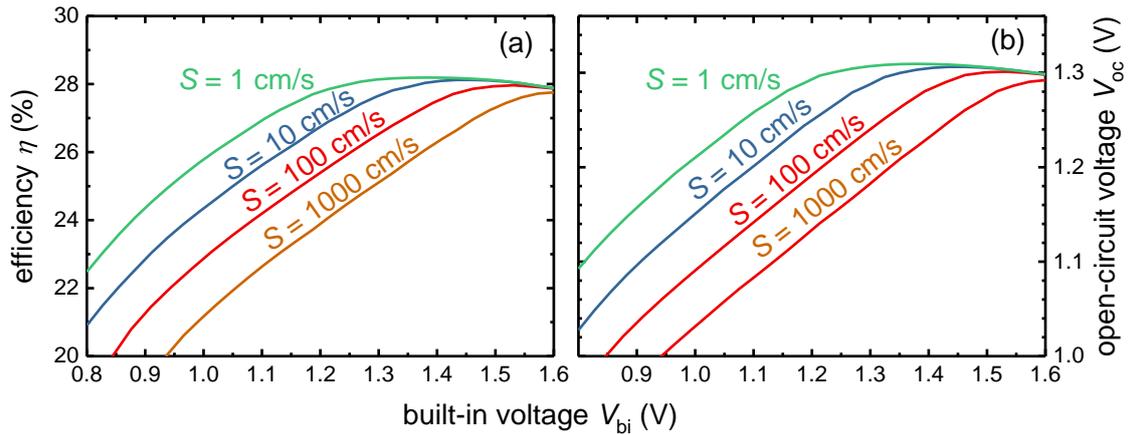

**Figure 8**: Result of drift-diffusion simulations of MAPI solar cells showing (a) power conversion efficiency and (b) open-circuit voltage $V_{oc}$ as a function of the built-in voltage for different values of the surface recombination velocity $S$ for minority carriers at front and back contact. For every value of $S$, there is an optimum built-in voltage, because for lower built-in voltages, surface recombination increases and for higher built-in voltages Auger recombination increases. (Online version in colour.)

Finally, an open question is the practical influence of the built-in voltage $V_{bi}$ on solar cell performance. Charge carrier diffusion lengths are generally reported to be high enough to achieve efficient charge collection via diffusion within absorbers with thicknesses of less than a micron. However, the built-in voltage still affects the concentration of minority carriers at the interfaces to electron and hole contact materials. Only a high $V_{bi}$ reduces the minority carrier concentration sufficiently to suppress recombination at these interfaces. Thus, for a given value of $S$, there is a minimum value for $V_{bi}$ that the cells need to have in order to not have additional losses at the interfaces. Figure 8 shows how high $V_{bi}$ has to be for different values of $S$ in order to achieve optimum efficiency. The simulations shown in Fig. 8 are done using the same method as used for Fig. 3, i.e. they include radiative recombination, photon recycling and Auger recombination. In the simulation, lower built-in voltages lead to increased surface recombination while higher built-in voltages lead to increased Auger recombination due to the higher concentration of majority carriers close to the contacts. Thus, there is always an optimum $V_{bi}$ for every value of $S$.

Figure 8 shows that while $V_{bi}$ can be smaller than $E_g/q$, it should be 'high enough' (which is defined differently depending on the values of $S$ as shown in Fig. 8) in order to reduce losses. However, it is often not clear whether the values of $V_{bi}$ leading to optimum power conversion efficiency in Fig. 8 are consistent with the expected workfunction difference between the contacts. For instance, for the high $V_{oc}$ device discussed in ref. [35], the electron and hole contact materials (PTAA and PCBM) are fairly intrinsic and the highly conductive materials ITO and Ag should not have a sufficiently high workfunction difference to enable efficient suppression of interfacial recombination. Nevertheless, the cells are in practice able to suppress recombination quite efficiently. This suggests that we lack knowledge on how the band diagram of the cell looks like and how the workfunctions and energy levels of the contact and absorber layers enable such low interfacial recombination *rates* (even if we do understand the low recombination *velocities*).

## Funding

The author acknowledges support from the Impuls- und Vernetzungsfonds der Helmholtz Gemeinschaft via the project PEROSEED and the Bavarian Ministry of Economic Affairs and Media, Energy and Technology for the joint projects in the framework of the Helmholtz Institute Erlangen-Nürnberg.

## References

(1) Green, M. A.; Hishikawa, Y.; Dunlop, E. D.; Levi, D. H.; Hohl-Ebinger, J.; Ho-Baillie, A. W. Y. 2018 Solar Cell Efficiency Tables (Version 52). *Prog. Photovolt: Res. Appl.*, **26**, 427-436. (doi: 10.1002/pip.3040)

(2) Yoshikawa, K.; Kawasaki, H.; Yoshida, W.; Irie, T.; Konishi, K.; Nakano, K.; Uto, T.; Adachi, D.; Kanematsu, M.; Uzu, H. et al. 2017 Silicon Heterojunction

Solar Cell With Interdigitated Back Contacts for a Photoconversion Efficiency Over 26%. *Nat. Energy*, **2**, 17032. (10.1038/nenergy.2017.32)

(3) Yoshikawa, K.; Yoshida, W.; Irie, T.; Kawasaki, H.; Konishi, K.; Ishibashi, H.; Asatani, T.; Adachi, D.; Kanematsu, M.; Uzu, H. et al. 2017 Exceeding Conversion Efficiency of 26% by Heterojunction Interdigitated Back Contact Solar Cell With Thin Film Si Technology. *Sol. Energy Mater. Sol. Cells*, **173**, 37-42. (10.1016/j.solmat.2017.06.024)

(4) Jackson, P.; Wuerz, R.; Hariskos, D.; Lotter, E.; Witte, W.; Powalla, M. 2016 Effects of Heavy Alkali Elements in $Cu(In,Ga)Se_2$ Solar Cells With Efficiencies Up to 22.6%. *Phys. Status Solidi RRL*, **10**, 583-586

(5) Friedlmeier, T. M.; Jackson, P.; Bauer, A.; Hariskos, D.; Kiowski, O.; Wuerz, R.; Powalla, M. 2015 Improved Photocurrent in Cu(In, Ga)Se-2 Solar Cells: From 20.8% to 21.7% Efficiency With CdS Buffer and 21.0% Cd-Free. *IEEE J. Photov.*, **5**, 1487-1491

(6) Jackson, P.; Hariskos, D.; Wuerz, R.; Kiowski, O.; Bauer, A.; Friedlmeier, T. M.; Powalla, M. 2015 Properties of $Cu(In,Ga)Se_2$ Solar Cells With New Record Efficiencies Up to 21.7%. *Phys Status Solidi-R*, **9**, 28-31

(7) Jeon, N. J.; Na, H.; Jung, E. H.; Yang, T. Y.; Lee, Y. G.; Kim, G.; Shin, H. W.; Il Seok, S.; Lee, J.; Seo, J. 2018 A Fluorene-Terminated Hole-Transporting Material for Highly Efficient and Stable Perovskite Solar Cells. *Nat. Energy*, **3**, 682-689

(8) Best Research Cell Efficiences https://www.nrel.gov/pv/assets/pdfs/pv-efficiency-chart.20181221.pdf. *https://www. nrel. gov/pv/assets/pdfs/pv-efficiency-chart. 20181221. pdf* **2018**.

(9) Li, H.; Xiao, Z.; Ding, L.; Wang, J. 2018 Thermostable Single-Junction Organic Solar Cells With a Power Conversion Efficiency of 14.62%. *Science Bulletin*, **63**, 340-342

(10) Kirchartz, T.; Rau, U. 2018 What Makes a Good Solar Cell? *Adv. Energy Mater.*, **8**, 1703385. (doi: 10.1002/aenm.201703385)

(11) Kirchartz, T.; Krückemeier, L.; Unger, E. L. 2018 Research Update: Recombination and Open-Circuit Voltage in Lead-Halide Perovskites. *Apl Materials*, **6**, 100702. (doi: 10.1063/1.5052164)

(12) De Wolf, S.; Holovsky, J.; Moon, S. J.; Löper, P.; Niesen, B.; Ledinsky, M.; Haug, F. J.; Yum, J. H.; Ballif, C. 2014 Organometallic Halide Perovskites: Sharp Optical Absorption Edge and Its Relation to Photovoltaic Performance. *J. Phys. Chem. Lett.*, **5**, 1035-1039. (doi: 10.1021/jz500279b)

(13) Blank, B.; Kirchartz, T.; Lany, S.; Rau, U. 2017 Selection Metric for Photovoltaic Materials Screening Based on Detailed-Balance Analysis. *Phys. Rev. Applied*, **8**, 024032

(14) Ahrenkiel, R. K. 1993 Minority-Carrier Lifetime in III-V Semiconductors. *Minority Carriers in Iii-V Semiconductors: Physics and Applications*, **39**, 39-150

(15) Abdi-Jalebi, M.; Andaji-Garmaroudi, Z.; Cacovich, S.; Stavrakas, C.; Philippe, B.; Richter, J. M.; Alsari, M.; Booker, E. P.; Hutter, E. M.; Pearson, A. J. et al. 2018 Maximizing and Stabilizing Luminescence From Halide Perovskites With Potassium Passivation. *Nature*, **555**, 497

(16) deQuilettes, D. W.; Zhang, W.; Burlakov, V. M.; Graham, D. J.; Leijtens, T.; Osherov, A.; Bulovic, V.; Snaith, H. J.; Ginger, D. S.; Stranks, S. D. 2016 Photo-Induced Halide Redistribution in Organic-Inorganic Perovskite Films. *Nat. Commun*, **7**, 11683. (Article)

(17) deQuilettes, D. W.; Vorpahl, S. M.; Stranks, S. D.; Nagaoka, H.; Eperon, G. E.; Ziffer, M. E.; Snaith, H. J.; Ginger, D. S. 2015 Impact of Microstructure on Local Carrier Lifetime in Perovskite Solar Cells. *Science*, **348**, 683-686

(18) Stranks, S. D.; Burlakov, V. M.; Leijtens, T.; Ball, J. M.; Goriely, A.; Snaith, H. J. 2014 Recombination Kinetics in Organic-Inorganic Perovskites: Excitons, Free Charge, and Subgap States. *Phys. Rev. Applied*, **2**, 034007. (10.1103/PhysRevApplied.2.034007)

(19) Stranks, S. D.; Eperon, G. E.; Grancini, G.; Menelaou, C.; Alcocer, M. J. P.; Leijtens, T.; Herz, L. M.; Petrozza, A.; Snaith, H. J. 2013 Electron-Hole Diffusion Lengths Exceeding 1 Micrometer in an Organometal Trihalide Perovskite Absorber. *Science*, **342**, 341-344

(20) Shockley, W.; Read, W. T. 1952 Statistics of the Recombination of Holes and Electrons. *Phys. Rev.*, **87**, 835-842

(21) Hall, R. N. 1952 Electron-Hole Recombination in Germanium. *Phys. Rev.*, **87**, 387

(22) Asbeck, P. 1977 Self-Absorption Effects on Radiative Lifetime in Gaas-Gaalas Double Heterostructures. *J. Appl. Phys.*, **48**, 820-822

(23) Staub, F.; Hempel, H.; Hebig, J. C.; Mock, J.; Paetzold, U. W.; Rau, U.; Unold, T.; Kirchartz, T. 2016 Beyond Bulk Lifetimes: Insights into Lead Halide Perovskite Films From Time-Resolved Photoluminescence. *Phys. Rev. Applied*, **6**, 044017

(24) Staub, F.; Kirchartz, T.; Bittkau, K.; Rau, U. 2017 Manipulating the Net Radiative Recombination Rate in Lead Halide Perovskite Films by Modification of Light Outcoupling. *The Journal of Physical Chemistry Letters*, **8**, 5084-5090. (doi: 10.1021/acs.jpclett.7b02224)

(25) Rau, U.; Paetzold, U. W.; Kirchartz, T. 2014 Thermodynamics of Light Management in Photovoltaic Devices. *Phys. Rev. B*, **90**, 035211

(26) Sproul, A. B. 1994 Dimensionless Solution of the Equation Describing the Effect of Surface Recombination on Carrier Decay in Semiconductors. *J. Appl. Phys.*, **76**, 2851-2854. (doi: 10.1063/1.357521)

(27) Casey, H. C.; Stern, F. 1976 Concentration-Dependent Absorption and Spontaneous Emission of Heavily Doped GaAs. *J. Appl. Phys.*, **47**, 631-643. (doi: 10.1063/1.322626)

(28) Stern, F. 1976 Calculated Spectral Dependence of Gain in Excited GaAs. *J. Appl. Phys.*, **47**, 5382-5386. (doi: 10.1063/1.322565)

(29) Werner, J. H.; Mattheis, J.; Rau, U. 2005 Efficiency Limitations of Polycrystalline Thin Film Solar Cells: Case of Cu(In,Ga)Se$_2$. *Thin Solid Films*, **480-481**, 399-409

(30) Trupke, T.; Green, M. A.; Würfel, P.; Altermatt, P. P.; Wang, A.; Zhao, J.; Corkish, R. 2003 Temperature Dependence of the Radiative Recombination Coefficient of Intrinsic Crystalline Silicon. *J. Appl. Phys.*, **94**, 4930-4937. (doi: 10.1063/1.1610231)

(31) Richter, A.; Glunz, S. W.; Werner, F.; Schmidt, J.; Cuevas, A. 2012 Improved Quantitative Description of Auger Recombination in Crystalline Silicon. *Phys. Rev. B*, **86**, 165202

(32) Staub, F.; Rau, U.; Kirchartz, T. 2018 Statistics of the Auger Recombination of Electrons and Holes Via Defect Levels in the Band Gap - Application to Lead-Halide Perovskites. *ACS Omega*, **3**, 8009-8016. (doi: 10.1021/acsomega.8b00962)

(33) Crothers, T. W.; Milot, R. L.; Patel, J. B.; Parrott, E. S.; Schlipf, J.; Müller-Buschbaum, P.; Johnston, M. B.; Herz, L. M. 2017 Photon Reabsorption Masks Intrinsic Bimolecular Charge-Carrier Recombination in CH$_3$NH$_3$PbI$_3$ Perovskite. *Nano Lett.*, **17**, 5782-5789. (doi: 10.1021/acs.nanolett.7b02834)

(34) Richter, J. M.; Abdi-Jalebi, M.; Sadhanala, A.; Tabachnyk, M.; Rivett, J. P. H.; Pazos-Outon, L. M.; Gödel, K. C.; Price, M.; Deschler, F.; Friend, R. H. 2016 Enhancing Photoluminescence Yields in Lead Halide Perovskites by Photon Recycling and Light Out-Coupling. *Nat. Commun.*, **7**, 13941. (Article)

(35) Liu, Z.; Krückemeier, L.; Krogmeier, B.; Klingebiel, B.; Marquez, J. A.; Levcenko, S.; Öz, S.; Mathur, S.; Rau, U.; Unold, T. et al. 2019 Open-Circuit Voltages Exceeding 1.26 V in Planar Methylammonium Lead Iodide Perovskite Solar Cells. *ACS Energy Lett.*, **4**, 110-117. (doi: 10.1021/acsenergylett.8b01906)

(36) Krogmeier, B.; Staub, F.; Grabowski, D.; Rau, U.; Kirchartz, T. 2018 Quantitative Analysis of the Transient Photoluminescence of CH$_3$NH$_3$PbI$_3$/PC$_{61}$BM Heterojunctions by Numerical Simulations. *Sustainable Energy Fuels*, **2**, 1027-1034. (10.1039/C7SE00603A)

(37) Olson, J. M.; Ahrenkiel, R. K.; Dunlavy, D. J.; Keyes, B.; Kibbler, A. E. 1989 Ultralow Recombination Velocity at Ga0.5In0.5P/GaAs Heterointerfaces. *Appl. Phys. Lett.*, **55**, 1208-1210. (doi: 10.1063/1.101656)

(38) Molenkamp, L. W.; van Blik, H. F. J. 1988 Very Low Interface Recombination Velocity in (Al,Ga)As Heterostructures Grown by Organometallic Vapor Phase Epitaxy. *J. Appl. Phys.*, **64**, 4253-4256. (doi: 10.1063/1.341298)

(39) Kuciauskas, D.; Kanevce, A.; Burst, J. M.; Duenow, J. N.; Dhere, R.; Albin, D. S.; Levi, D. H.; Ahrenkiel, R. K. 2013 Minority Carrier Lifetime Analysis in the Bulk of Thin-Film Absorbers Using Subbandgap (Two-Photon) Excitation. *IEEE J. Photov.*, **3**, 1319-1324. (10.1109/JPHOTOV.2013.2270354)

(40) Zhao, Y.; Boccard, M.; Liu, S.; Becker, J.; Zhao, X. H.; Campbell, C. M.; Suarez, E.; Lassise, M. B.; Holman, Z.; Zhang, Y. H. 2016 Monocrystalline CdTe Solar Cells With Open-Circuit Voltage Over 1V and Efficiency of 17%. *Nat. Energy*, **1**, 16067. (10.1038/nenergy.2016.67)

(41) Brendel, R.; Peibst, R. 2016 Contact Selectivity and Efficiency in Crystalline Silicon Photovoltaics. *IEEE J. Photov.*, **6**, 1413-1420

(42) Roe, E. T.; Egelhofer, K. E.; Lonergan, M. C. 2018 Limits of Contact Selectivity/Recombination on the Open-Circuit Voltage of a Photovoltaic. *ACS Appl. Energy Mater.*, **1**, 1037-1046. (doi: 10.1021/acsaem.7b00179)

(43) Courtier, N. E.; Cave, J. M.; Foster, J. M.; Walker, A. B.; Richardson, G. 2019 How Transport Layer Properties Affect Perovskite Solar Cell Performance: Insights From a Coupled Charge Transport/Ion Migration Model. *Energ. Environ. Sci.*, **12**, 396-409. (10.1039/C8EE01576G)

(44) Tessler, N.; Vaynzof, Y. 2018 Preventing Hysteresis in Perovskite Solar Cells by Undoped Charge Blocking Layers. *ACS Appl. Energy Mater.*, **1**, 676-683. (doi: 10.1021/acsaem.7b00176)

(45) Kirchartz, T.; Bisquert, J.; Mora-Sero, I.; Garcia-Belmonte, G. 2015 Classification of Solar Cells According to Mechanisms of Charge Separation and Charge Collection. *Phys. Chem. Chem. Phys.*, **17**, 4007-4014

(46) Braly, I. L.; deQuilettes, D. W.; Pazos-Outon, L. M.; Burke, S.; Ziffer, M. E.; Ginger, D. S.; Hillhouse, H. W. 2018 Hybrid Perovskite Films Approaching the Radiative Limit With Over 90% Photoluminescence Quantum Efficiency. *Nat. Photon.*, **12**, 355-361

(47) Mattheis, J.; Werner, J. H.; Rau, U. 2008 Finite Mobility Effects on the Radiative Efficiency Limit of Pn-Junction Solar Cells. *Phys. Rev. B*, **77**, 085203

(48) Mattheis, J. Mobility and homogeneity effects on the power conversion efficiency of solar cells. Universität Stuttgart, p. 140, Jan 2008.

(49) Kirchartz, T.; Staub, F.; Rau, U. 2016 Impact of Photon Recycling on the Open-Circuit Voltage of Metal Halide Perovskite Solar Cells. *ACS Energy Lett.*, **1**, 731-739. (doi: 10.1021/acsenergylett.6b00223)

(50) Wang, J.; Fu, W.; Jariwala, S.; Sinha, I.; Jen, A. K. Y.; Ginger, D. S. 2018 Reducing Surface Recombination Velocities at the Electrical Contacts Will Improve Perovskite Photovoltaics. *ACS Energy Lett.*, **4**, 222-227. (doi: 10.1021/acsenergylett.8b02058)

(51) Taretto, K.; Soldera, M.; Troviano, M. 2013 Accurate Explicit Equations for the Fill Factor of Real Solar Cells - Applications to Thin-Film Solar Cells. *Prog. Photovolt: Res. Appl.*, **21**, 1489-1498. (doi: 10.1002/pip.2235)

(52) Jacobs, D. A.; Catchpole, K. R.; Beck, F. J.; White, T. P. 2016 A Re-Evaluation of Transparent Conductor Requirements for Thin-Film Solar Cells. *J. Mater. Chem. A*, **4**, 4490-4496. (10.1039/C6TA01670G)

(53) Sendner, M.; Nayak, P. K.; Egger, D. A.; Beck, S.; Muller, C.; Epding, B.; Kowalsky, W.; Kronik, L.; Snaith, H. J.; Pucci, A. et al. 2016 Optical Phonons in Methylammonium Lead Halide Perovskites and Implications for Charge Transport. *Mater. Horiz.*, **3**, 613-620. (10.1039/C6MH00275G)

(54) Calado, P.; Telford, A. M.; Bryant, D.; Li, X.; Nelson, J.; O'Regan, B. C.; Barnes, P. R. F. 2016 Evidence for Ion Migration in Hybrid Perovskite Solar Cells With Minimal Hysteresis. *Nat. Commun.*, **7**, 13831. (Article)

(55) Belisle, R. A.; Nguyen, W. H.; Bowring, A. R.; Calado, P.; Li, X.; Irvine, S. J. C.; McGehee, M. D.; Barnes, P. R. F.; O'Regan, B. C. 2017 Interpretation of Inverted Photocurrent Transients in Organic Lead Halide Perovskite Solar Cells: Proof of the Field Screening by Mobile Ions and Determination of the Space Charge Layer Widths. *Energ. Environ. Sci.*, **10**, 192-204. (10.1039/C6EE02914K)

(56) Turren-Cruz, S. H.; Hagfeldt, A.; Saliba, M. 2018 Methylammonium-Free, High-Performance, and Stable Perovskite Solar Cells on a Planar Architecture. *Science*, **362**, 449. (10.1126/science.aat3583)

(57) Shockley, W.; Queisser, H. J. 1961 Detailed Balance Limit of Efficiency of Pn-Junction Solar Cells. *J. Appl. Phys.*, **32**, 510-519

(58) Nayak, P. K.; Cahen, D. 2013 Updated Assessment of Possibilities and Limits for Solar Cells. *Adv. Mater.*, **26**, 1622-1628. (doi: 10.1002/adma.201304620)

(59) Polman, A.; Knight, M.; Garnett, E. C.; Ehrler, B.; Sinke, W. C. 2016 Photovoltaic Materials: Present Efficiencies and Future Challenges. *Science*, **352**. (10.1126/science.aad4424)

(60) Kirchartz, T.; Kaienburg, P.; Baran, D. 2018 Figures of Merit Guiding Research on Organic Solar Cells. *The Journal of Physical Chemistry C*, **122**, 5829-5843. (doi: 10.1021/acs.jpcc.8b01598)

(61) Luo, D.; Yang, W.; Wang, Z.; Sadhanala, A.; Hu, Q.; Su, R.; Shivanna, R.; Trindade, G. F.; Watts, J. F.; Xu, Z. et al. 2018 Enhanced Photovoltage for Inverted Planar Heterojunction Perovskite Solar Cells. *Science*, **360**, 1442. (10.1126/science.aap9282)

(62) Stolterfoht, M.; Wolff, C. M.; Marquez, J. A.; Zhang, S.; Hages, C. J.; Rothhardt, D.; Albrecht, S.; Burn, P. L.; Meredith, P.; Unold, T. et al. 2018 Visualization and Suppression of Interfacial Recombination for High-Efficiency Large-Area Pin Perovskite Solar Cells. *Nat. Energy*, **3**, 847-854. (10.1038/s41560-018-0219-8)

(63) Saidaminov, M. I.; Kim, J.; Jain, A.; Quintero-Bermudez, R.; Tan, H.; Long, G.; Tan, F.; Johnston, A.; Zhao, Y.; Voznyy, O. et al. 2018 Suppression of Atomic Vacancies Via Incorporation of Isovalent Small Ions to Increase the Stability of Halide Perovskite Solar Cells in Ambient Air. *Nat. Energy*,

(64) Zheng, X.; Chen, B.; Dai, J.; Fang, Y.; Bai, Y.; Lin, Y.; Wei, H.; Zeng, X.; Huang, J. 2017 Defect Passivation in Hybrid Perovskite Solar Cells Using Quaternary Ammonium Halide Anions and -Cations. *Nat. Energy*, **2**, 17102. (10.1038/nenergy.2017.102)

(65) Wang, Z.; Lin, Q.; Chmiel, F. P.; Sakai, N.; Herz, L. M.; Snaith, H. J. 2017 Efficient Ambient-Air-Stable Solar Cells With 2D-3D Heterostructured Butylammonium-Caesium-Formamidinium Lead Halide Perovskites. *Nat. Energy*, **2**, 17135. (10.1038/nenergy.2017.135)

(66) Saliba, M.; Matsui, T.; Domanski, K.; Seo, J. Y.; Ummadisingu, A.; Zakeeruddin, S. M.; Correa-Baena, J. P.; Tress, W. R.; Abate, A.; Hagfeldt, A. et al. 2016 Incorporation of Rubidium Cations into Perovskite Solar Cells Improves Photovoltaic Performance. *Science*, **354**, 206. (10.1126/science.aah5557)

(67) Tan, H.; Jain, A.; Voznyy, O.; Lan, X.; Garcia de Arquer, F. P.; Fan, J. Z.; Quintero-Bermudez, R.; Yuan, M.; Zhang, B.; Zhao, Y. et al. 2017 Efficient and Stable Solution-Processed Planar Perovskite Solar Cells Via Contact Passivation. *Science*, **355**, 722-726. (10.1126/science.aai9081)

(68) Yang, W. S.; Park, B. W.; Jung, E. H.; Jeon, N. J.; Kim, Y. C.; Lee, D. U.; Shin, S. S.; Seo, J.; Kim, E. K.; Noh, J. H. et al. 2017 Iodide Management in Formamidinium-Lead-Halide-Based Perovskite Layers for Efficient Solar Cells. *Science*, **356**, 1376. (10.1126/science.aan2301)

(69) Hou, Y.; Du, X.; Scheiner, S.; McMeekin, D. P.; Wang, Z.; Li, N.; Killian, M. S.; Chen, H.; Richter, M.; Levchuk, I. et al. 2017 A Generic Interface to Reduce the Efficiency-Stability-Cost Gap of Perovskite Solar Cells. *Science*, **358**, 1192. (10.1126/science.aao5561)

(70) Anaraki, E. H.; Kermanpur, A.; Steier, L.; Domanski, K.; Matsui, T.; Tress, W.; Saliba, M.; Abate, A.; Grätzel, M.; Hagfeldt, A. et al. 2016 Highly Efficient and Stable Planar Perovskite Solar Cells by Solution-Processed Tin Oxide. *Energ. Environ. Sci.*, **9**, 3128-3134. (10.1039/C6EE02390H)

(71) Green, M. A. 1981 Solar Cell Fill Factors: General Graph and Empirical Expressions. *Solid State Electron*, **24**, 788-789

(72) Green, M. A. 1982 Accuracy of Analytical Expressions for Solar-Cell Fill Factors. *Solar Cells*, **7**, 337-340

(73) Miyata, K.; Meggiolaro, D.; Trinh, M. T.; Joshi, P. P.; Mosconi, E.; Jones, S. C.; De Angelis, F.; Zhu, X. Y. 2017 Large Polarons in Lead Halide Perovskites. *Sci Adv*, **3**. (10.1126/sciadv.1701217)

(74) Emin, D. 2018 Barrier to Recombination of Oppositely Charged Large Polarons. *J. Appl. Phys.*, **123**, 055105. (doi: 10.1063/1.5019834)

(75) Tong, C. J.; Li, L.; Liu, L. M.; Prezhdo, O. V. 2018 Long Carrier Lifetimes in PbI2-Rich Perovskites Rationalized by Ab Initio Nonadiabatic Molecular Dynamics. *ACS Energy Lett.*, 1868-1874. (doi: 10.1021/acsenergylett.8b00961)

(76) Long, R.; Prezhdo, O. V. 2015 Dopants Control Electron–Hole Recombination at Perovskite–TiO$_2$ Interfaces: Ab Initio Time-Domain Study. *Acs Nano*, **9**, 11143-11155. (doi: 10.1021/acsnano.5b05843)

(77) Azarhoosh, P.; McKechnie, S.; Frost, J. M.; Walsh, A.; van Schilfgaarde, M. 2016 Research Update: Relativistic Origin of Slow Electron-Hole Recombination in Hybrid Halide Perovskite Solar Cells. *Apl Materials*, **4**, 091501. (doi: 10.1063/1.4955028)

(78) Ambrosio, F.; Wiktor, J.; De Angelis, F.; Pasquarello, A. 2018 Origin of Low Electron-Hole Recombination Rate in Metal Halide Perovskites. *Energ. Environ. Sci.*, **11**, 101-105. (10.1039/C7EE01981E)

(79) Meggiolaro, D.; Motti, S. G.; Mosconi, E.; Barker, A. J.; Ball, J.; Andrea Riccardo Perini, C.; Deschler, F.; Petrozza, A.; De Angelis, F. 2018 Iodine Chemistry Determines the Defect Tolerance of Lead-Halide Perovskites. *Energ. Environ. Sci.*, **11**, 702-713. (10.1039/C8EE00124C)

(80) Hutter, E. M.; Gelvez-Rueda, M. C.; Osherov, A.; Bulovic, V.; Grozema, F. C.; Stranks, S. D.; Savenije, T. J. 2017 Direct-Indirect Character of the Bandgap in Methylammonium Lead Iodide Perovskite. *Nat. Mater.*, **16**, 115-120. (10.1038/nmat4765)

(81) Wang, T.; Daiber, B.; Frost, J. M.; Mann, S. A.; Garnett, E. C.; Walsh, A.; Ehrler, B. 2017 Indirect to Direct Bandgap Transition in Methylammonium Lead Halide Perovskite. *Energ. Environ. Sci.*, **10**, 509-515. (10.1039/C6EE03474H)

(82) Zhang, X.; Shen, J. X.; Wang, W.; Van de Walle, C. G. 2018 First-Principles Analysis of Radiative Recombination in Lead-Halide Perovskites. *ACS Energy Lett.*, 2329-2334. (doi: 10.1021/acsenergylett.8b01297)

(83) Kirchartz, T.; Rau, U. 2017 Decreasing Radiative Recombination Coefficients Via an Indirect Band Gap in Lead Halide Perovskites. *The Journal of Physical Chemistry Letters*, **8**, 1265-1271. (doi: 10.1021/acs.jpclett.7b00236)

(84) Kirchartz, T.; Markvart, T.; Rau, U.; Egger, D. A. 2018 Impact of Small Phonon Energies on the Charge-Carrier Lifetimes in Metal-Halide Perovskites. *The Journal of Physical Chemistry Letters*, **9**, 939-946. (doi: 10.1021/acs.jpclett.7b03414)

(85) Markvart, T. 1981 Semiclassical Theory of Non-Radiative Transitions. *Journal of Physics C: Solid State Physics*, **14**, L895

(86) Ridley, B. K. 1978 On the Multiphonon Capture Rate in Semiconductors. *Solid State Electron*, **21**, 1319-1323

(87) Markvart, T. Multiphonon recombination. In *Recombination in Semiconductors*, Landsberg, P. T., Ed.; Cambridge University Press: Cambridge, 2003; pp 467-468.

(88) Ridley, B. K. *Quantum Processes in Semiconductors;* Oxford University Press: Oxford, 2013.

(89) Markvart, T. Multiphonon recombination. In *Recombination in Semiconductors*, Landsberg, P. T., Ed.; Cambridge University Press: Cambridge, 2003; p 475.

(90) Du, M. H. 2015 Density Functional Calculations of Native Defects in $CH_3NH_3PbI_3$: Effects of Spin-Orbit Coupling and Self-Interaction Error. *The Journal of Physical Chemistry Letters*, **6**, 1461-1466. (doi: 10.1021/acs.jpclett.5b00199)

(91) Uratani, H.; Yamashita, K. 2017 Charge Carrier Trapping at Surface Defects of Perovskite Solar Cell Absorbers: A First-Principles Study. *The Journal of Physical Chemistry Letters*, **8**, 742-746. (doi: 10.1021/acs.jpclett.7b00055)

(92) Mayers, M. Z.; Tan, L. Z.; Egger, D. A.; Rappe, A. M.; Reichman, D. R. 2018 How Lattice and Charge Fluctuations Control Carrier Dynamics in Halide Perovskites. *Nano Lett.*, **18**, 8041-8046. (doi: 10.1021/acs.nanolett.8b04276)

(93) Benduhn, J.; Tvingstedt, K.; Piersimoni, F.; Ullbrich, S.; Fan, Y.; Tropiano, M.; McGarry, K. A.; Zeika, O.; Riede, M. K.; Douglas, C. J. et al. 2017 Intrinsic Non-Radiative Voltage Losses in Fullerene-Based Organic Solar Cells. *Nat. Energy*, **2**, 17053. (10.1038/nenergy.2017.53)

(94) Azzouzi, M.; Yan, J.; Kirchartz, T.; Liu, K.; Wang, J.; Wu, H.; Nelson, J. 2018 Nonradiative Energy Losses in Bulk-Heterojunction Organic Photovoltaics. *Phys. Rev. X*, **8**, 031055. (10.1103/PhysRevX.8.031055)

(95) Liu, X.; Li, Y.; Ding, K.; Forrest, S. 2019 Energy Loss in Organic Photovoltaics: Nonfullerene Versus Fullerene Acceptors. *Phys. Rev. Applied*, **11**, 024060. (10.1103/PhysRevApplied.11.024060)

(96) Stolterfoht, M.; Caprioglio, P.; Wolff, C. M.; Marquez, J. A.; Nordmann, J.; Zhang, S.; Rothhardt, D.; Hörmann, U.; Redinger, A.; Kegelmann, L.; Albrecht, S.; Kirchartz, T.; Saliba, M.; Unold, T.; Neher, D. The perovskite/transport layer interfaces dominate non-radiative recombination in efficient perovskite solar cells. Unpublished Work, 2018.

(97) Aberle, A. G.; Wenham, S. R.; Green, M. A. A new method for accurate measurements of the lumped series resistance of solar cells. 1993; pp 133-139.

(98) Pysch, D.; Mette, A.; Glunz, S. W. 2007 A Review and Comparison of Different Methods to Determine the Series Resistance of Solar Cells. *Sol. Energy Mater. Sol. Cells*, **91**, 1698-1706

(99) Müller, T. C. M.; Pieters, B. E.; Rau, U.; Kirchartz, T. 2013 Analysis of the Series Resistance in Pin-Type Thin-Film Silicon Solar Cells. *J. Appl. Phys.*, **113**

(100) Helbig, A.; Kirchartz, T.; Schaeffler, R.; Werner, J. H.; Rau, U. 2010 Quantitative Electroluminescence Analysis of Resistive Losses in Cu(In, Ga)Se(2) Thin-Film Modules. *Sol. Energy Mater. Sol. Cells*, **94**, 979-984